\documentclass[%
 reprint,
 amsmath,amssymb,
 aps,
]{revtex4-1}

\usepackage{graphicx}
\usepackage{dcolumn}
\usepackage{bm}
\usepackage[dvipsnames]{xcolor}
\definecolor{linkcolor}{rgb}{0.18, 0.1875, 0.5725}

\usepackage{hyperref}
\hypersetup{
	colorlinks=true,
	citecolor=linkcolor,
	linkcolor=linkcolor,
	urlcolor=linkcolor
}
\usepackage[mathlines]{lineno}

\usepackage{natbib}
\begin{document}
\preprint{APS/123-QED}

\title{Prospect of using Grover's search in the noisy-intermediate-scale quantum-computer era}

\author{Yulun Wang}

\author{Predrag S. Krstic}
\email[]{krsticps@gmail.com} 
\affiliation{Institute for Advanced Computational Science, Stony Brook University, Stony Brook NY 11794-5250, USA}

\date{\today}

\begin{abstract}
In order to understand the bounds of utilization of the Grover’s search algorithm for the large unstructured data in presence of the quantum computer noise, we undertake a series of simulations by inflicting various types of noise, modelled by the IBM QISKit. We apply three forms of Grover’s algorithms: (1) the standard one, with 4-10 qubits, (2) recently published modified Grover’s algorithm, set to reduce the circuit depth, and (3) the algorithms in (1) and (2) with multi-control Toffoli’s modified by addition of an ancilla qubit. Based on these simulations, we find the upper bound of noise for these cases, establish its dependence on the quantum depth of the circuit and provide comparison among them. By extrapolation of the fitted thresholds, we predict what would be the typical gate error bounds when apply the Grover’s algorithms for the search of a data in a data set as large as thirty two thousands. 
\end{abstract}

\maketitle
\section{\label{sec:intro}INTRODUCTION\protect}
Grover’s algorithm (GA) \cite{grover97,boyer98} for search of unstructured data shows obvious and convincing quantum advantage to the classical search algorithms. Thus, GA scales the number of search iterations (i.e. the search time) with $\sqrt{N}$ rather than with N in the classical search, where N is the number of searched data. This polynomial acceleration has been proved to be optimal for the data search problems \cite{Zalka1999}: A search job that would run on classical computer one month would take about 3.5 hours to finish on a quantum computer using Grover’s algorithm. 

\begin{figure}
\includegraphics{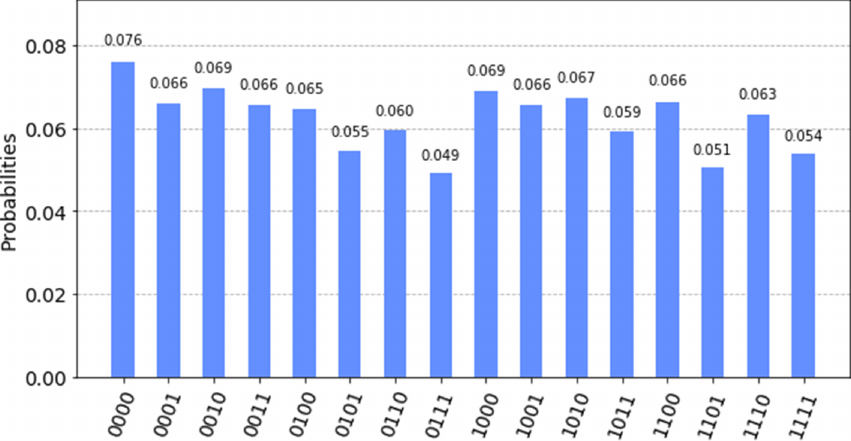}
\caption{\label{fig:1} The 4-qubit Grover search of \texttt{0011} in a set of 16 data on IBM’s latest quantum computer "\texttt{ibmq\_paris}". The ideal probability to measure “\texttt{0011}” is 96.7\%, however the measured one is only 6.6\% due to the quantum noise and the specific topology of the machine.}
\end{figure}

Running algorithms such as Grover’s search on the modern, quantum-circuit based quantum computers is achieved by consecutive unitary operations. However, the imperfect quantum gates and thermally induced decoherence in the NISQ (Noisy-Intermediate-Scale-Quantum) computers are the major source of noise in current hardware, producing errors in the quantum operations \cite{preskill18}. Size of quantum circuit is characterized by the total number of gates, and often by the circuit depth, the largest number of gates along any input-output path, moving forward in time. Thus, circuit depth is proportional to the smallest amount of time steps to execute the circuit assuming that each gate is performed within a time step, and the gates that acts on independent qubits can be executed simultaneously. Many researchers have analyzed the impact of various noise types in Grover’s search algorithm \cite{reitzner19,rao12,pablo99,shapira03,salas08}. With a large circuit depth, quantum program for complex tasks propagate and accumulate errors throughout the whole quantum circuits. As a consequence, the search for the targeted data fails because of the small signal-to-noise ratio. For example, with the current level of noise in the superconducting quantum devices (like are IBM Q and Rigetti), one could clearly select an element among 8 data (3 qubits), while search of an element among the 16 data (4 qubits) fails. We show in Fig.~\ref{fig:1} the results of the search of \texttt{0011} element at the 4-qubits Hilbert space using an IBM Q computer of the latest generation. Due to the noise the targeted state probability is not distinguishable from the probabilities of the other states. A deterministic version of GA \cite{Long2001} has been proposed to keep the ideal target probability 100\% for any size of database, which is slightly improved from the standard GA (SGA) \cite{grover97}. But it does not necessarily help to distinguish a searched target in presence of the noise. For the SGA, the circuit depth and total number of gates exponentially increase with the number of qubits, which might induce exponential magnification of the noise assuming the worst case that errors are produced with some probability at each gate. For example, circuit depth with 3 qubits is 58 (95 gates), with 4 qubits the depth grows to 242 (322 gates), while for 6 qubits it reaches 1922 (2418 gates). 

In this work we define the selectivity $S$ by 
\begin{equation}
S=10\log_{10} {(P_t/P_{hn})}
\end{equation}
to be the metrics that quantifies the performance of the quantum algorithm, where $P_t$ is the measured probability of the targeted state and $P_{hn}$ is probability of the highest measured noise signal. The selectivity decreases with increase of the noise probabilities. In Fig.~\ref{fig:1}, the largest noise-induced probability is of the \texttt{0000} state (7.6\%) which is higher than the target probability (6.6\%), leading to a negative selectivity, indicating an unsuccessful search. In this study we chose $S=3$ for the lowest acceptable threshold of the selectivity, which corresponds to the $P_t/P_{hn}\approx2$. When $S=3$ is reached, the item searched by the Grover’s algorithm is considered well distinguishable from the noise, and we consider that algorithm is successfully executed. Thus, this value of the selectivity defines the highest acceptable bound of the noise for the successful Grover’s operation. The threshold error probabilities and damping parameters in this work are obtained by interpolating the computed data to the selectivity $S=3$.

The goal of this paper is not to mitigate the errors in Grover’s quantum circuit, but rather to predict how small these errors need to be in order to reach the acceptable selectivity of the targeted state. This is done by replicating various types of noise in the circuits and performing simulation of the Grover’s algorithm by varying number of qubits n from 4 to 10, which alters the size of the searched data sets as $2^n$. The quantum circuit depth is increased exponentially with number of qubits using SGA reflecting the dependence on n of Multi-Control Toffoli (MCT) gates present in oracle and diffusion operator as well as the number of iterations, proportional to $2^{n/2}$ (see Note SI in the Supplemental Material). However, this increase can be restrained by adding ancillas to MCTs (labelled here with MCTA) in both the SGA and the recently published Grover’s algorithms modified (MGA) for reduction of the circuit depth \cite{zhang20}. With fewer gates in a circuit, smaller gate operational errors are accumulated, and the coherent time domain is increased. We find that adding one ancilla to the MCTs as done in IBM Quantum Information Science Kit (QISKit) \cite{Qiskit} following Barenco et al. \cite{barenco95}, strongly reduces the circuit depth and consequently reduces overall noise in all studied cases.

In Section~\ref{sec:2} we introduced the background of noise simulation, as defined in QISKit. In Sections~\ref{sec:3} and~\ref{sec:4} we perform experiments of simulation for various errors by varying the number of qubits (i.e. varying the number of gates in the quantum circuit) to obtain the upper bound of the thresholds toward a successful Grover’s search. SGA with use of MCTs and MCTAs are studied in Section~\ref{sec:3} for the circuit depth and response to the various errors in the circuit. The same is done in Section~\ref{sec:4} studying error response of modified GA which achieve circuit depth reduction by use of the local diffusion operators \cite{zhang20}. In Section~\ref{sec:5} we provide comparison of the selectivity thresholds due to the errors between all considered algorithms, highlighting the most successful ones. Finally, in Section~\ref{sec:6} we give our conclusions.

\section{\label{sec:2}Noise Simulation\protect}
Various types of errors have been identified and characterized in superconducting quantum computers. The design of a quantum computer can be evaluated by the coherence time of a qubit (natural relaxation time, $T_1$, and the time for the qubit dephasing from the superposition state, $T_2$), as well as by the gate errors (single-qubit rotations and two-qubit operations, such as CNOT) \cite{tannu19} and readout errors. Soft methods have been developed to mitigate particular error types. However, the utilization of these methods makes the quantum circuit more complex, bringing in more gates and more errors, which limits their effectiveness. 

In the real quantum devices, qubits suffer simultaneously from various types of the gate errors and decoherence simultaneously, which makes extraction of the effects of the individual errors a formidable task. The error probabilities are often characterized by the qubit multiplicity of the various quantum operations. For the latest superconducting quantum-computers from IBM (for example, \texttt{ibmq\_cambridge} at 18:23:30 on May 4th, 2020), single qubit instructions have an average error probability of 0.093\% for $U_2$ gate operations and 0.19\% for $U_3$ gate operations, while the CNOT gates have an average error probability of 3.5\%. The $U_3$ gate is defined in QISKit as single-qubit rotation gate with three Euler angles: $\theta$, $\phi$ and $\lambda$ by $U_3(\theta,\phi,\lambda)=R_Z(\phi)R_X(-\frac{\pi}{2})R_Z(\theta)R_X(\frac{\pi}{2})R_X(\lambda)$ where $R_Z(\theta)$ and $R_X(\theta)$ are the single-qubit rotations around Z and X axis with an angle of $\theta$ on the Bloch sphere, while $U_2$ gate is defined as $U_2(\phi,\lambda)=U_3(\frac{\pi}{2},\phi,\lambda)$. Besides gate errors, this machine also suffers from a significant readout error at average 9.5\% and limited coherent time (average $T_1$ = 82 $\mu$s and $T_2$ = 41 $\mu$s). 

We study the effects of the gate and coherent qubit time errors as modelled in the QISKit (version 1.16.0) \cite{Qiskit}. Gate errors such as bit and phase flip (so called Pauli errors) and depolarization are modelled by assuming that an error happens with some probability in each gate in the quantum circuit \cite{knill08}. The Kraus operators for the Pauli errors are defined in QISKit as: $E_0=\sqrt{1-p_\sigma}I$, $E_1=\sqrt{p_\sigma}\sigma$, where $p_\sigma$ is the error probability and $\sigma=X$ (bit flip), $Z$ (phase flip), $Y$ (combined flips), respectively \cite{Qiskit}. These operators yield the mixed states $\rho'=\varepsilon(\rho)=E_0\rho E_0^\dagger+E_1\rho E_1^\dagger=(1-p_\sigma)\rho+p_\sigma\sigma\rho\sigma$, where the initial state is described by density operator $\rho$ and the functional $\varepsilon$ is the quantum operation. Qubit is flipped with probability $p_\sigma$ or left unchanged with probability $1-p_\sigma$. With depolarization error a qubit is depolarized with a probability $p_{dep}$. The depolarization channel is defined as $\varepsilon(\rho)=(1-p_{dep})\rho+p_{dep}\frac{I}{2^n}$, where $I$ is completely mixed state density and $n$ is the number of qubits in the error channel \cite{nielsen}.

Keeping a quantum state coherent is essential for implementation of gate operations and measurements of qubits in quantum computing. Induced by external disturbances, decoherence erodes the fidelity of quantum executions due to coupling between quantum system and environment. Thus, decoherence of quantum states is large obstacle toward scalable quantum computing. We study the Grover’s circuit response to the amplitude damping (AD) and phase damping (PD). In these cases the dimensionless damping probability $p_{dam}$ defines the Kraus operators $E_{0,1}$ \cite{nielsen,gutierrez13} as $E_0\vert0\rangle=\vert0\rangle, E_0\vert1\rangle=\sqrt{1-p_{dam}}\vert1\rangle, E_1\vert0\rangle=0$ for both AD and PD, while $E_1\vert1\rangle=\sqrt{p_{dam}}\vert x\rangle$, where $x=0$ for AD and $x=1$ for PD, $p_{dam}$ is the probability of either the energy loss (AD) or information loss (PD) to the environment in a small time interval $\delta t$. After these dissipation processes repeat many ($m$) times in succession, during the gate time $t_g=m\delta t$, the corresponding transition channel decays exponentially \cite{preskilllecture18} as $e^{-\Gamma t_g}$, where $\Gamma$ is the probability rate $\Gamma=\frac{p_{dam}}{\delta t}$. The amplitude damping is often referred as relaxation process for the time $T_1$, where $T_1=\frac{1}{\Gamma}$, and $\Gamma$ is the amplitude damping probability rate. Similarly, the phase damping is referred as the superposition dephasing process for the time $T_2$, where $T_2=\frac{1}{\Gamma}$ and $\Gamma$ is the phase damping rate. $T_1$ and $T_2$ together are employed to characterize the lifetime of the qubit amplitude and phase. The quantum error channels mentioned above are applied to all single qubit quantum operations during the noise simulation. The error channels for two qubit operation are obtained by applying the single qubit error to each of two qubits (for example, CNOT) \cite{Qiskit}. 

QISKit supports simulation of the thermal relaxation mode by inputting values for $T_1$ and $T_2$ with pre-defined gate time $t_g$, where $T_2 \leq 2T_1$\cite{Grifoni99,yafet1963}. The gate time $t_g$ in the model is set to 50 ns for single qubit $U_2$ rotations, 100 $n_s$ for $U_3$ rotations, 300 ns for CNOT gates, 1000 ns for qubit reset and 1000 ns for measurements. The $T_1$ and $T_2$ relaxation error rates are defined as $\varepsilon_{T_1}=e^{-{t_g}/{T_1}}$ and $\varepsilon_{T_2}=e^{-{t_g}/{T_2}}$, respectively \cite{blank20}. 

We also apply our error analysis of the Grover’s algorithm (with MCTs or MCTAs) to the method targeting to lower the circuit depth (and fewer number of gates), recently published by Zhang et al \cite{zhang20}. They achieved reduction of the number of gates by replacing the standard Grover operator with the depth modified one in adjusted sequences, in which standard diffusion operator is replaced by local diffusion operator \cite{younes13}. Furthermore, the depth-reduced algorithm can be executed in multi-stages to eliminate noise. For example, the search target can be divided in two stages: $\vert t\rangle=\vert t_1\rangle\vert t_2\rangle$. After the first stage $\vert t_1\rangle$, partial measurement is applied to terminate search in part of qubits. The quantum measurements in first stage can avoid qubit idling with unwanted noise in the second stage. The rest of qubits are reset and reinitialized before the execution of the second stage, which also eliminates the unwanted noise accumulated in the first stage. We here apply MCTA gates of this modified Grover’s algorithm, obtaining further reduction of the circuit depth, thus increasing the thresholds in all studied quantum errors.

\section{\label{sec:3}GROVER’S CIRCUIT WITH ANCILLARY QUBITS\protect}

MCT gates are the key components of oracle and diffusion operators for the Grover’s search. Current IBM quantum computers allow only elementary executions such as single qubit rotations and controlled two qubit gates. Thus, the multiple qubits Toffoli gates are decomposed to elementary operations during circuit transpiling by QISKit and hence occupy the biggest part of the circuit depth, resulting in main sources of noise. By using MCTAs the circuit depth and number of total gates can be dramatically reduced. In this work, non-ancilla Toffoli gates and 1-ancilla Toffoli gates are used for composing circuits with different depth. Circuits for the MCT and MCTA gates are generated by QISKit functions (using \texttt{qiskit.aqua.circuits.gates.mct} with mode “\texttt{noancilla}” and “\texttt{advanced}” which are based on grey-code sequence and recursive splitting method for non-ancilla and 1-ancilla Toffoli gates respectively \cite{barenco95}). When designing quantum circuits, various searched targets have their unique oracles. Only one MCT gate is required to construct oracle for the encoded search state $\vert1\rangle$. Hence for the simplest circuit, $\vert1\rangle^{\otimes n}$is chosen as searched target item in all tests \cite{figgatt17}. The gate numbers and circuit depth are calculated with QISKit functions by varying the number of qubits from 4 to 14, for the SGA as well as for the circuits where MCTAs replaced MCTs (SGAA). These are shown in the Table SI of the Supplemental Material. For example, in n=4 qubits SGA 3 iterations (marked as $G_4^3$) are applied, which yields 322 for the total number of gates. When using MCTAs, the reduction of the number of gates for n=4 is about 1.3 times. In 10-qubit algorithm, the number of gates in the circuit is reduced nearly 10 times by using MCTAs. For 14-qubits algorithm this reduction is nearly 50. This leads to a noticeable degradation of noise, especially when increasing number of qubits, as our results below show.

\begin{figure}
\includegraphics{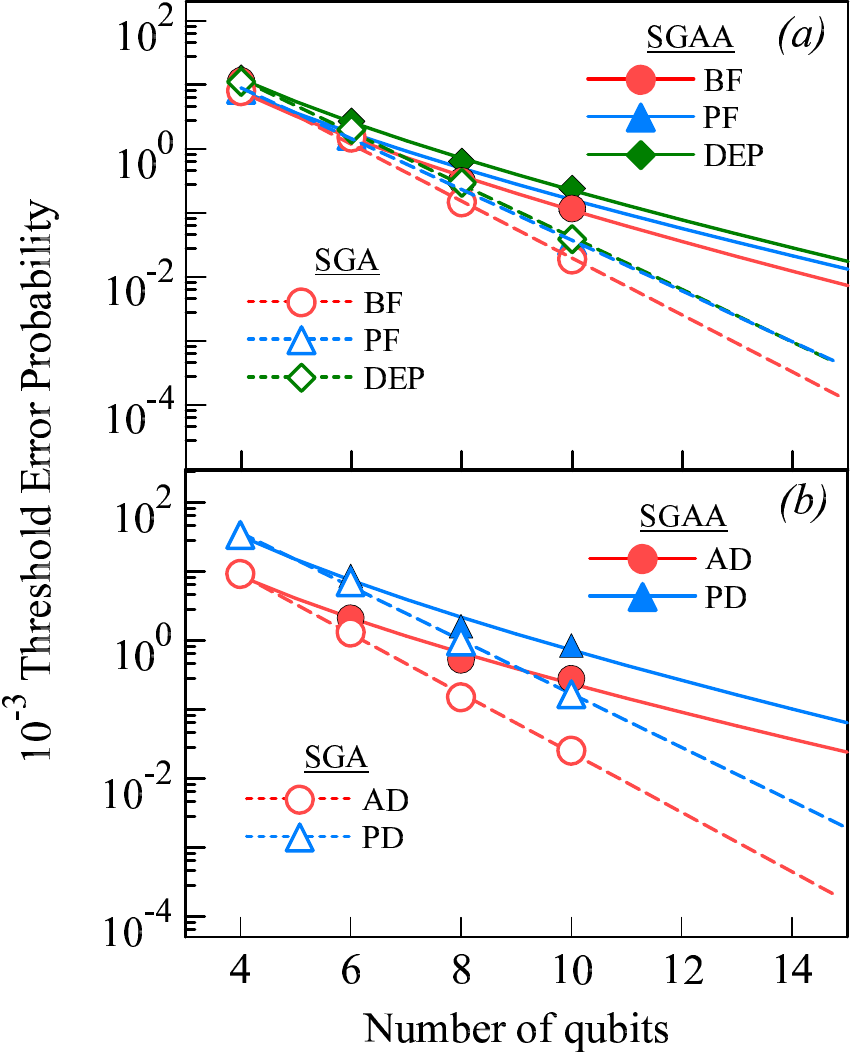}
\caption{\label{fig:2} Threshold error probabilities for (a) bit flip (BF), phase flip (PF) and depolarizing (DEP), and (b) threshold damping parameters of amplitude damping (AD) and phase damping (PD), with various number of qubits in GA search. The hollow symbols and solid symbols are for SGA and SGAA, respectively. Dashed (SGA) and solid (SGAA) lines are fitting curves with extrapolation, as described in the text.}
\end{figure}

We calculated the upper bounds of noise defined by the lowest acceptable selectivity defined in the Introduction. We present these bounds in Fig.~\ref{fig:2} for both SGA and SGAA, by changing the types of noise and varying number of qubits from 4 to 10. As discussed in the Introduction, using MCTAs in SGA increases the upper bound of thresholds in all studied cases as a consequence of significant contraction of the quantum depths and the total number of gates (see Table SI). This contraction increases with number of qubits. Even though the gate errors are accumulated after each gate operation, noise is significantly suppressed with the reduction of the number of gates. From Fig.~\ref{fig:2}, circuit for 4-qubit SGA requires the Pauli, depolarization and amplitude damping error probabilities below $10^{-2}$ and phase damping below $3\times10^{-2}$ for acceptable search result, with similar conclusions for SGAA. With 8 qubits search, SGA and SGAA start deviating from each other. The 10 qubit SGA circuit requires the error probabilities below $10^{-5}$ (except for phase damping $<10^{-4}$), while an order of magnitude lower circuit depth of SGAA than SGA produces a leap in selectivity i.e. error thresholds are of the order of $10^{-4}$.Thus, when searching a database of size of $2^{10}$, the upper bound of depolarization error, for example, is about $2\times10^{-4}$. Similar conclusions can be obtained for amplitude ($<2\times10^{-4}$) and phase ($<7\times10^{-4}$) damping noise, as well as for bit-flip and phase flip error probabilities ($<10^{-4}$): the threshold parameter values increase an order of magnitude in 10 qubits case when MCTAs are used. 

The selectivity thresholds in Fig.~\ref{fig:2} are directly correlated to the number of gates in the circuits when varying the number of qubits. To understand this correlation, we first fitted the data for the number of gates in Table SI of Supplemental Material vs. number of qubits (4 to 14 qubits) using the functional forms inspired by Barenco at al. \cite{barenco95} analysis of the number of gates in the MCTs with and without ancilla. This is explained in detail in Note SI of the Supplemental Material. For SGA case, the number of gates ($G$) is an exponential function of $n$, with fitting function $G_{SGA}=4.6991e^{1.0388n}$, reflecting the  number of iterations, $2^{n/2}$, as well as $2^{n}$ functional dependence of MCTs on $n$. However, combined product of exponential and power dependence fits best SGAA cases, $G_{SGAA}=1.2761n^{2.8401}e^{0.3436n}$. The exponent here corresponds to the $2^{n/2}$ number of iterations, which confirms the polynomial dependence of the number of gates on a MCT with ancilla \cite{barenco95}. Similar functional dependences are obtained for the circuit depths of SGA and SGAA, due to the almost constant ratio of number of gates and the circuit depth in Table SI. Since it is expected that the quantum errors are accumulated proportionally to the number of gates in a circuit, it is not surprising that the best fits for the selectivity thresholds (lines in Fig.~\ref{fig:2}) are obtained with the similar fitting models as for $G_{SGA}$ and $G_{SGAA}$, even with similar values of exponents. The fitted curves, obtained with high correlations, are shown in Fig.~\ref{fig:2} with dashed and solid lines and fitting parameters are listed in Table SII of the Supplemental Material. Although the simulations were done with the top-of-the-line supercomputing cluster \cite{seawulf}, 10-qubits was the limit for successful simulation of GA with QISKit, mainly due to the large memory requirements in a classical computer. Still, by extrapolation of the obtained fits we can predict with certainty the selectivity thresholds for larger number of qubits. Thus, for SGA with 15 qubits the threshold error probabilities are as low as $10^{-6}$ for phase damping and $10^{-7}$ for other types of errors. However, the SGAA improves the error threshold by about 2 orders of magnitude. For example, the upper bound of depolarizing noise increases from about $4\times10^{-7}$ to $1.75\times10^{-5}$.

\begin{figure}
\includegraphics{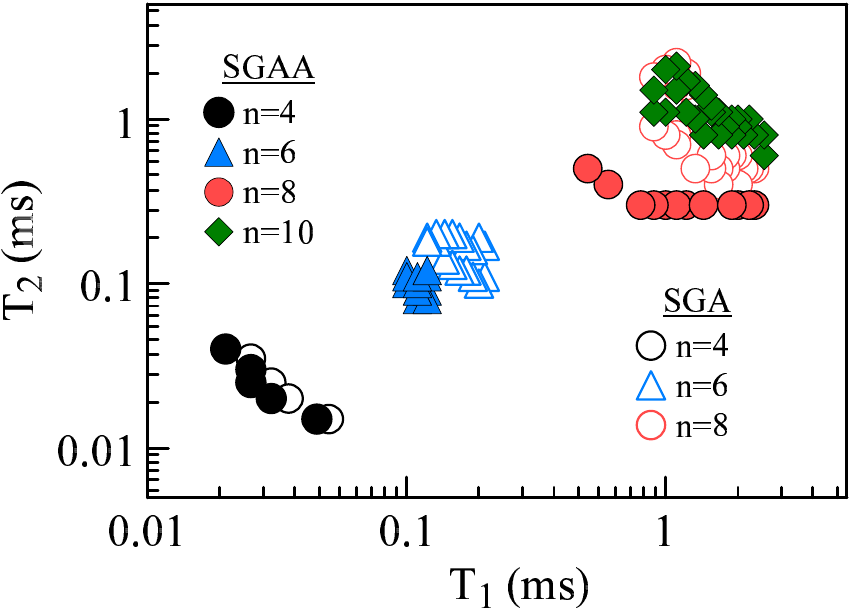}
\caption{\label{fig:3} Selectivity thresholds for the thermal relaxation times of SGA and SGAA with various number of qubits.}
\end{figure}

Calculated variations of the thermal relaxation times $T_1$ and $T_2$ with number of qubits are presented in Fig.~\ref{fig:3}. Symbols in the figure show thresholds of the lower bound for $T_1$ and $T_2$ defined for the selectivity in range 2.5 to 3.5 (i.e. $P_{t}/P_{hn}$ in range 1.778 to 2.239). Unlike the other types of noise where it was favorable to have a bigger threshold, $T_1$ and $T_2$ are desired to have lower thresholds which leads to a smaller coherent time for execution of SGA and SGAA algorithms. Each point in the figure is independently calculated and collected for given $T_1$ and $T_2$. The results for the 10-qubit SGA search could not be obtained because the threshold $T_1$ and $T_2$ exceed the QISKit simulation limits (2,500 microseconds). For example, successful search through a set of $2^4$ data can be done with $T_1$ and $T_2$ in the range 15-50 $\mu$s with both SGA and SGAA, which is well achievable by current quantum hardware. For $2^6$ dataset, the thresholds for $T_1$ and $T_2$ increase to about 100 $\mu$s with SGAA and close to 200 $\mu$s with SGA, while for the 8-qubit search $T_1$ and $T_2$ average around 1150 $\mu$s and 330 $\mu$s, respectively with SGAA, but about 1230 $\mu$s and 590 $\mu$s , respectively, with SGA. For 10-qubit SGA, the requirements for $T_1$ and $T_2$ are exceeding 10 ms with a total circuit depth of 128002. Better results are obtained with SGAA. Thus, the relaxation time limits for $T_1$ and $T_2$ with 10 qubits average to 1.8 and 1 ms, respectively.

\section{\label{sec:4}GROVER’S CIRCUIT BY LOCAL DIFFUSION OPERATORS\protect}

The selectivity thresholds are calculated in this section for the GA with modified circuit to reduce the quantum depth \cite{zhang20} (MGA) in two variants, one-stage (M1GA) and two-stage (M2GA) with inclusion of noise, as was done in Section~\ref{sec:3}. In addition, we also studied MGA with one ancilla in MCTs (MkGAA, k=1,2), which showed, like in case of SGAA, a further improvement in the error thresholds. Table SIII in Supplementary Material contains studied configurations as well as the information on the relevant quantum depths and number of gates (including both one- and two-qubit gates). One-stage and two-stage methods of depth optimization are both tested for 4-10 qubit cases. Fig. S2 in Note SII of the Supplemental Material illustrates schematically, as an example, the quantum circuits for SGA and MGA configurations with 4 qubits. The MGA circuit configurations in Table SIII are also studied using MCTAs in place of MCTs, convincingly reducing the depth of the circuits.

\begin{figure}
\includegraphics{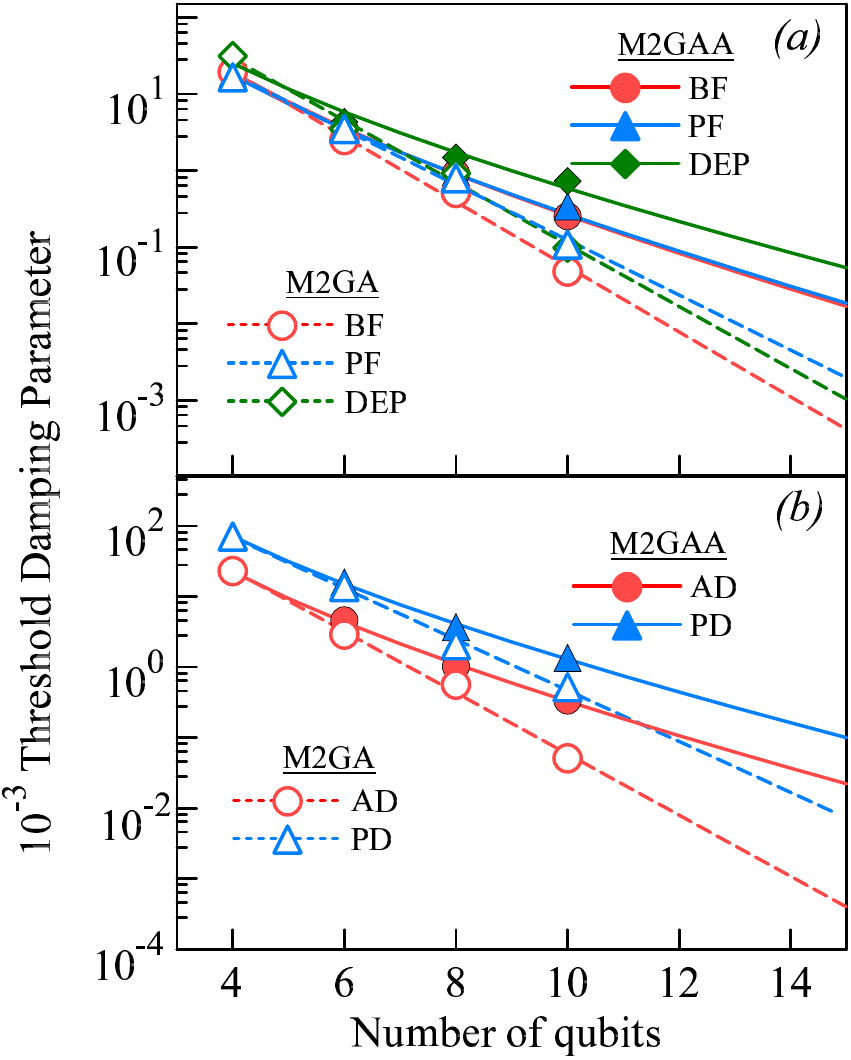}
\caption{\label{fig:4} Fitting curve of selectivity thresholds for the error probability vs. number of qubits for various types of errors (BF, PF and DEP in (a) and AD and PD in (b)) applied at two-stage depth reduced GA \cite{zhang20} with use of MCTs (M2GA) and of MCTAs(M2GAA). Fig. S3 in the Supplemental Material shows the calculated data for the one-stage depth reduced GA.}
\end{figure}

The dependencies of the number of gates on the number of qubits have the functional forms used in Section~\ref{sec:3} and discussed in Note SI of the Supplemental Material. As one would expect, the numbers of gates for M1GA and M2GA fit well the exponential function of the number of qubits, $G_{M1GA}=5.0785e^{0.9953n}$ and $G_{M2GA}=3.8835e^{1.0220n}$, respectively. On the other hand, power-times-exponential function fits well number of gates vs. number of qubits for M1GAA and M2GAA, i.e. $G_{M1GAA}=1.2823n^{2.5439}e^{0.3880n}$ and $G_{M2GAA}=0.7670n^{2.7057}e^{0.4125n}$. The extrapolations of the fitting functions imply that M1GA circuit requires over 15 million gates with the 15 qubits and nearly 18 million gates for M2GA. Comparing with 27 million gates for SGA in Section~\ref{sec:3}, the improvement of M1GA and M2GA is significant though might not be sufficient for practical applications. However, using MCTAs these numbers drastically decreases to about 424k and 568k for M1GAA and M2GAA, respectively, comparable of 484k in case of SGAA. Thus, one could expect a huge reduction of noise in the circuits using MCTAs, as was already shown in Section~\ref{sec:3} for the SGA. 

Having in mind correlation of the number of gates and the selection thresholds for the error probabilities, found in Section~\ref{sec:3}, we apply the functional forms used for the number of gates to fit the error thresholds versus number of qubits for the MGA circuits. Tables SIV and SV in the Supplemental Material show the fitting coefficients in all cases. The calculated data for the selectivity thresholds as well as the fitting curves are shown in Fig.~\ref{fig:4} and extrapolated to 15 qubits. Like in the SGA cases of Fig.~\ref{fig:2}, by setting the threshold of errors to $10^{-4}$ it follows that M2GA can perform a successful search with datasets of up to $2^8$ size. Similar results are obtained for M1GA and presented in Fig. S3 of the Supplemental Material. Most error threshold upper bounds are below that limits with 10 qubits search. However, with the use of MCTAs this data size limit extends to $2^{12}$ and $2^{14}$ for M1GAA and M2GAA, respectively. 

\begin{figure}
\includegraphics{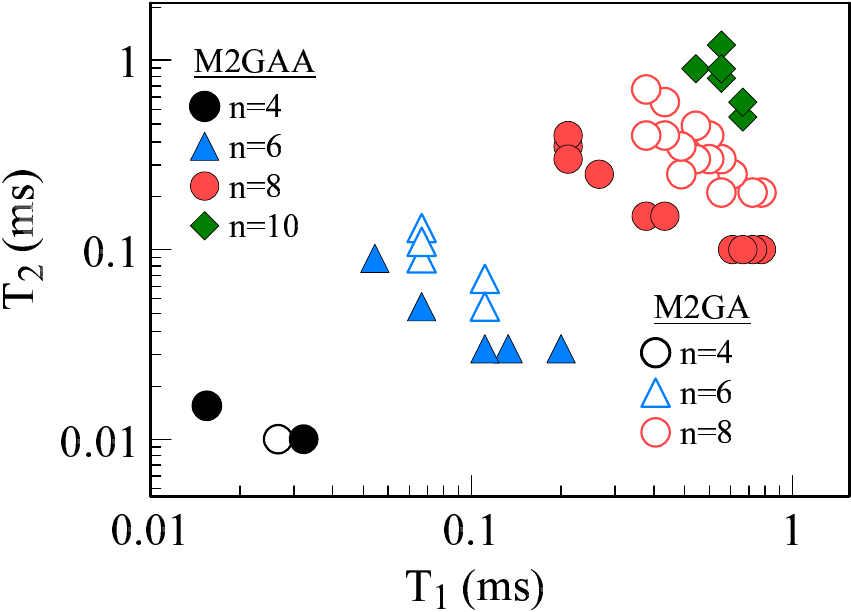}
\caption{\label{fig:5} Selectivity thresholds for thermal noise of two-stage modified Grover’s search \cite{zhang20}, (M2GA and M2GAA). Similar results calculated for the one-stage algorithm are shown in Fig. S4 of the Supplemental Material.}
\end{figure}

We also investigate the effect of decoherence by tuning $T_1$ and $T_2$, like in Section~\ref{sec:3}. Data points with a selectivity in the range 2.5 to 3.5 are collected and plotted in Fig.~\ref{fig:5} for two-stage depth-reduced method \cite{zhang20}. Similar results for M1GA and M1GAA are calculated and shown in Fig. S4 of the Supplemental material. The thresholds $T_1$ and $T_2$ for 4-qubit are as low as 15 and 15 $\mu$s with M2GA. But for 8 qubits with M2GA, the averages of $T_1$ and $T_2$ decrease to 500 $\mu$s and 350 $\mu$s from 1159 $\mu$s and 330 $\mu$s with SGA. Use of MCTAs helps to further reduce the threshold relaxation times. The coherent requirements for the selectivity with M2GAA yield average values for $T_1$ and $T_2$ of 110 and 45 $\mu$s, respectively (6 qubits) and 400 $\mu$s for $T_1$ and 200 $\mu$s for $T_2$ (8 qubits). For 10 qubit, successful GA can be achieved by M2GAA with average of 600 and 800 $\mu$s for $T_1$ and $T_2$, respectively, which is about 3 times shorter for $T_1$ and about 20\% for $T_2$ from SGAA in Section~\ref{sec:3}.

\section{\label{sec:5}COMPARISON AND DISCUSSION OF THE RESULTS\protect}

\begin{figure}
\includegraphics{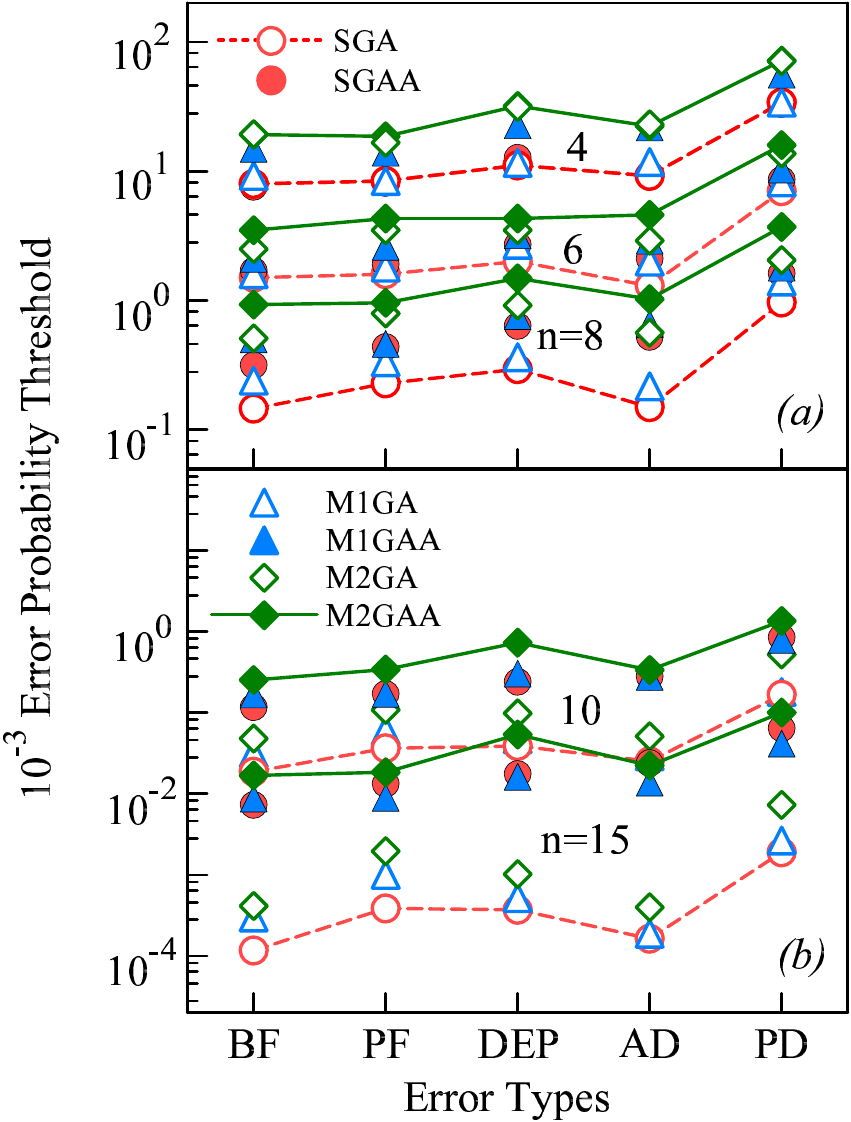}
\caption{\label{fig:6} Comparison of the selectivity thresholds due to various quantum circuits errors in standard GA (SGA and SGAA when MCTA are used), one-stage (M1GA and M1GAA) and two-stage (M2GA and M2GAA) depth reduced GA for (a) 4,6 and 8, and (b) 10 and 15 qubits. The 15-qubit case was obtained by extrapolation of the fitting curves in Sections~\ref{sec:3} and~\ref{sec:4}. Higher resolution figures are presented in Fig. S6 of the Supplemental Material.}
\end{figure}

Comparisons of the selectivity thresholds for the error probabilities of the various types of error, obtained with algorithms in Sections~\ref{sec:3} and~\ref{sec:4} for 4-10 qubits as well as for the extrapolations to 15 qubits are presented at Figs.~\ref{fig:6}(a) (4-8 qubits) and ~\ref{fig:6}(b) (10 and 15 qubits). Due to the presence of quantum noise, measured probability of the targeted state for large number of qubits is at a low value. However, the targeted data is still selective and can be distinguished from other data. For example, in the 10-qubit M1GA with depolarizing noise (Fig. S5), probabilities of the targeted data are about factor 2 larger than the highest noise signal, and about 2 orders of magnitude higher than most of the noise signals, which ensures a successful Grover’s search. With 4 qubits, SGA, SGAA and M1GA have very similar error thresholds with a slightly bigger values of M1GA due to the depth reduction by using local diffusion operators. By applying MCTA in M1GA the selectivity thresholds are increased 1.5 (BF and PF) to 2 times (DEP, AD, PD). This is not a case for M2GA when n=4, which has similar error threshold values with M2GAA for all types, though slightly bigger than M1GAA. However, these relations for the selectivity thresholds are not kept when increasing the number of qubits. Thus, for 10 qubits SGAA is significantly more selective than both M1GA (10 times for AD, about 5 times for others) and M2GA, and is quite close to the values of M1GAA for all error types. Only M2GAA is convincingly most selective, leading by about factor 2 over SGAA and M1GAA except for AD and PD error types where it is only slightly better. The trend of separation of the algorithms which use MCTAs from the algorithms with MCTs is continuing with further increase of the number of qubits by extrapolation of the fitting curves in Sections~\ref{sec:3} and~\ref{sec:4} (and Tables SII, SIV and SV of the Supporting Material). Thus, with 15 qubits this separation reaches 2 orders of magnitude. In that case, the M2GA is up to a factor 2 more selective than M1GA and 2-4 times more selective than SGA. The trend for relations of M2GAA with M1GAA and SGAA is similar with 15 as it is with 10 qubits. 

Among all gate errors, depolarization has the highest error threshold in all configurations and is expected to have the least impact to the selectivity of the results in GA search. Similarly, the phase damping error thresholds are significantly bigger than these with amplitude damping in all cases. Concerning the considered algorithms, the best results are obtained by two-stage algorithms, whether these are used with MCTs or MCTAs. Moreover, with MCTAs used in the quantum circuits, the number of gates decrease with number of qubits changes from exponential dependence to the product of power and a weak exponential dependence. This leads to a huge improvement in the selectivity loss due to the noise in the circuit, qualifying M2GAA but also SGAA and M1GAA for a successful Grover’s search with 10 qubits if error probabilities are smaller than $10^{-4}$, i.e. with 15 qubits when the error probabilities are smaller than $10^{-5}$.

Therefore, using MCTAs in the quantum circuit in place of MCTs dramatically improves the noise resistance. Using MCTAs in GA is as effective as use of local diffusion operators in reducing the effect of gate noise, obtained in \cite{zhang20}. From discussion in Section~\ref{sec:4} it follows also that the two stage MGA contributes the most in reduction of decoherence.

\begin{figure}
\includegraphics{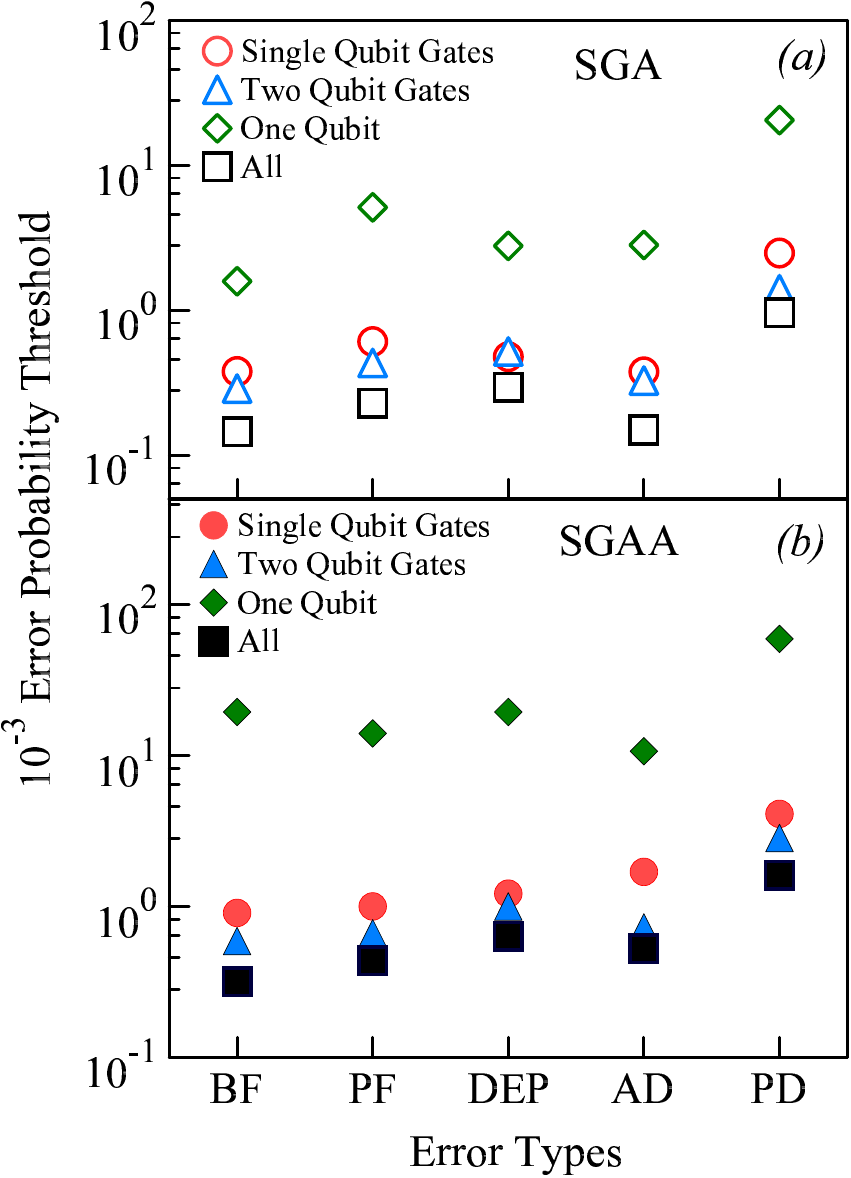}
\caption{\label{fig:7} Error threshold of BF, PF, DEP, AD and PD with noise channels applied only on single qubit gates and two qubit gates for 8-qubit (a) SGA and (b) SGAA. Simulated results with only one noisy qubit are also shown, as well as contribution of all error channels.}
\end{figure}

Nevertheless, the two-stage depth optimization is not supported on current IBM quantum computers because no quantum operations are allowed in the circuits after a measurement (See Fig. S2(c)  in Note SII of the Supplemental Material). Another impediment for implementing the optimal configurations is that all simulations in this paper are based on the ideal assumption that all qubits are fully connected to each other, which is not true for the current superconducting quantum computers. Extra SWAP gates could be added to circuit for adapting to the actual device topology which can dramatically increase the circuit depth after circuit transpiling to basis operations. Other computing platforms could be more successful in handling the quantum noise and thus implementing successfully larger number of qubits in Grover’s search in the near future. For example, the trapped ion quantum computers are recognized to have an exceptionally long coherence time, very high fidelity of gate operations, state preparation and readout with a high fidelity and fully connectivity of qubits \cite{Bruzewicz2019}. Besides, the symmetry features of a trapped ion device allow for more compact version of Grover search with smaller quantum depth. \cite{Ivanov2010}

Another drawback in our work could be the way we applied the noise models: These are applied to all qubits in quantum operations in GA. This seems like overestimation of the noise reality. However, this approach could partially compensate for our treatment of the two-qubit gates (2QGs). It is experimentally known that superconducting two-qubit gates have significantly higher error probability than one-qubit gates. In our model an error after a 2QG simply arises from the tensor product of the qubit’s states, which is likely underestimating the error of the 2QGs. For example, the number of 2QGs in both the 8-qubit SGA and SGAA circuits is very close to that of one-qubit gates. We show this by including only noisy two-qubit gates in the SGA (Fig.~\ref{fig:7}(a)) and SGAA (Fig.~\ref{fig:7}(b)). We get slightly lower error threshold than when including only noisy single qubit gates, indicating that in our model two-qubit operations contribute only slightly more to a final noise of the GA circuits. Another simplifying assumption is that all qubits are equally susceptible to errors, which is certainly not true in the real superconducting computers. We also test application of noise to only one qubit. This gives, in 8-qubit case about an order of magnitude for SGA (20-60 times for SGAA) higher error threshold than all qubits GA operations.

\section{\label{sec:6}CONCLUSIONS\protect}

We undertake a series of computer simulations of the Grover search by applying the noise, modelled in the IBM QISKit. We apply three forms of Grover's algorithms: (1) the standard one, with 4-10 qubits, (2) recently published modified Grover’s algorithm \cite{zhang20}, set to reduce the circuit depth, and (3) the algorithms in (1) and (2) with multi-control Toffoli’s modified by addition of 1 ancilla qubit (MCTAs). The noise/errors included are the bit and phase flips, depolarization, amplitude and phase damping, as well as the energy and phase relaxation times, determining the system coherence time. The circuits with MCTAs in all cases show a significant improvement of the selectivity thresholds for the error probabilities, which goes up to one order of magnitude for 10-qubit algorithm, and even more for larger number of qubits. This is explained by the exponential growth with number of qubits $n$ when MTAs are used, which transforms into combination of a power law and weak exponential growth, when MCTAs are utilized. These result into similar functional dependences on $n$ (with flipped sign of both the exponents and powers) for the selectivity thresholds due to the errors. The depth modified Grover’s algorithm shows increase of the error thresholds and decrease of threshold relaxation times, which are also notable improved by the use of MCTAs. By extrapolation of the fitted functional dependences to $n$ as large as 15, we also provide predictions of the error thresholds for successful search with all studied quantum circuit configurations, which set the limit for errors probabilities to $10^{-5}$ for successful search of database as large as thirty two thousand. While these errors might be beyond anticipated hardware possibilities, the error limit of $10^{-4}$ seems to be applicable in the near future for a GA search with 10 qubits, i.e. for a data set as large as 1000.

\begin{acknowledgments}
The authors acknowledge financial support from the Institute for Advanced Computational Science at Stony Brook University. We would like to thank Stony Brook Research Computing and Cyberinfrastructure, and the Institute for Advanced Computational Science at Stony Brook University for access to the high-performance SeaWulf computing system, which was made possible by National Science Foundation grant (\#1531492). The authors are grateful to Kun Zhang and Vladimir Korepin for the inspiring discussions.
\end{acknowledgments}

\bibliography{ms-revisedf}

\end{document}